\documentclass[twocolumn,showpacs,amsmath,amssymb,nofootinbib,superscriptaddress]{revtex4}

\usepackage{graphicx}
\usepackage{epsfig}
\usepackage{bm}
\usepackage{amsfonts}
\usepackage{amssymb}
\usepackage{color}
\usepackage{float}
\usepackage{amsmath}                  
\usepackage{dcolumn}
\usepackage{hyperref}
\usepackage{array}
\usepackage{lscape}
\usepackage{ulem}
\usepackage{xcolor}

\usepackage{graphicx,subfigure,bm,color,psfrag,hyperref}

\usepackage{mathtools}

\usepackage{amsmath,amsthm}
\usepackage{amsfonts}
\usepackage{amssymb}
\usepackage{verbatim}

\newcommand{\lin}{\\[7pt]}

\newcommand{\pder}[2]{\dfrac{\partial#1}{\partial#2}}

\newcommand{\de}{\mathrm{d}}

\begin{document}

\title{Cosmology of Lorentz fiber-bundle induced scalar-tensor theories}
 
\author{S. Ikeda}
\affiliation{Ikeda Institute of Applied Geometry, Ashimori 762-3, Okayama 701-1463, Japan}

\author{Emmanuel N. Saridakis}
\email{msaridak@phys.uoa.gr}
\affiliation{Department of Physics, National Technical University of Athens, Zografou
Campus GR 157 73, Athens, Greece}
\affiliation{National Observatory of Athens, Lofos Nymfon, 11852 Athens, 
Greece}
 \affiliation{Department of Astronomy, School of Physical Sciences, University 
of Science 
and Technology of China, Hefei 230026, P.R. China}

\author{P. C. Stavrinos}
\email{pstavrin@math.uoa.gr}
\affiliation{  Department of Mathematics, National and Kapodistrian University of Athens,
Panepistimiopolis 15784, Athens, Greece}
 
\author{A. Triantafyllopoulos}
\email{alktrian@phys.uoa.gr}
\affiliation{Section of Astrophysics, Astronomy and Mechanics, Department of Physics,  
National and Kapodistrian University of Athens, Panepistimiopolis 15784,
Athens, Greece}

\begin{abstract}
We investigate the cosmological applications of scalar-tensor theories that arise 
effectively from the Lorentz fiber bundle of a Finsler-like geometry. We first show that 
the involved nonlinear connection induces a new scalar degree of freedom and eventually a 
scalar-tensor theory. Using both a holonomic and a non-holonomic basis, we show the 
appearance of an effective dark energy sector, which additionally acquires 
an explicit interaction with the matter sector, arising purely from the internal 
structure 
of the theory. Applying the theory at late times we find that we can obtain the 
thermal history of the universe, namely the sequence of matter and dark-energy 
epochs, 
and moreover  the effective dark-energy equation-of-state parameter can be 
quintessence-like, phantom-like, or experience the phantom-divide crossing during 
cosmological evolution. Furthermore, applying the scenario at early times we see 
that one can acquire an exponential de Sitter solution, as well as obtain an inflationary 
realization with the desired scale-factor evolution. These features   arise 
purely from the intrinsic geometrical structure of Finsler-like geometry,  and reveal the 
capabilities of the construction.

\end{abstract}

\pacs{98.80.-k, 95.36.+x, 04.50.Kd}
\keywords{cosmology, geometry:Finsler-like, modified gravity}

\maketitle

\section{Introduction}

According to an increasing amount of  data the universe experienced two 
phases of accelerating expansion, one at early and one at late cosmological times. Such a 
behavior may imply a form of extension of  our current knowledge, in order to introduce 
the necessary extra degrees of freedom that would be needed for its successful 
explanation. In principle there are two main ways one could follow. The first is to 
assume that general relativity is correct however the matter content of the universe 
should be modified through the introduction of
the inflaton  
\cite{Bartolo:2004if} and/or dark energy fields \cite{Copeland:2006wr,Cai:2009zp}. The 
second way is to consider that the gravitational theory is not general relativity but a 
more fundamental theory, possessing the former as a particular limit, but which in 
general can provide the extra degrees of freedom needed for a successful cosmological 
description \cite{Capozziello:2011et,Nojiri:2010wj}).

In order to construct gravitational modifications one starts from the Einstein-Hilbert 
Lagrangian of general relativity and  includes extra terms, such as in
 $f(R)$ gravity 
\cite{Starobinsky:1980te,Capozziello:2002rd}
and  Lovelock
gravity \cite{Lovelock:1971yv, Deruelle:1989fj}, or even he can use torsion such as in 
$f(T)$ 
gravity  
\cite{Cai:2015emx} and in $f(T,T_G)$ gravity
\cite{Kofinas:2014owa}.
Alternatively, one may consider the general class of scalar-tensor theories, which 
include one extra scalar degree of freedom with general couplings with curvature terms. 
One general family of scalar-tensor theories is the Horndeski construction 
\cite{Horndeski:1974wa}, re-discovered in the framework of generalized galileons 
\cite{Nicolis:2008in,Deffayet:2009mn}.
 
One interesting class of modified gravity theories may arise through the more radical 
modification of the underlying geometry itself, namely considering Finsler or 
Finsler-like geometries \cite{kour-stath-st 
2012,Triantafyllopoulos:2018bli,Minas:2019urp,Basilakos:2013hua,Basilakos:2013ij,
Kouretsis:2010vs,Mavromatos:2010jt}. These geometries extend in a natural way the 
Riemannian one, by allowing the physical quantities to have a dependence on the 
observer 4-velocity, which in turn reflects the Lorentz-violating character of the 
kinematics 
\cite{Bogoslovsky:1999pp,Chang:2007vq,Vacaru:2010fi,Kostelecky:2008be,
Foster:2015yta,Kostelecky:2011qz,Kostelecky:2012ac,Stavrinos:2012ty,
Hohmann:2016pyt,Hohmann:2018rpp,Pfeifer:2019wus}. Furthermore, since 
Finsler and Finsler-like geometries are strongly related to the effective 
geometry within anisotropic media \cite{Born}, entering naturally   the analogue 
gravity approach \cite{Barcelo:2005fc}, they may play an important role in quantum 
gravity considerations. In such a geometrical setup, the dependence of the metric and 
other quantities on both the position coordinates of the 
base-manifold as well as on the directional/velocity variables of the tangent space or 
scalar/spinor variables, makes the tangent bundle, or a fiber bundle of a smooth 
manifold, 
the natural geometrical framework for their description. Finally, the Riemannian case is 
reproduced from Finsler geometry if 
the velocity-dependence is set to zero.

If one applies the above into a cosmological framework he obtains Finsler and 
Finsler-like 
cosmologies. The 
basic feature of these is the appearance of extra terms in the  Friedmann equations 
due to the intrinsic geometrical spacetime   anisotropy of Finsler and Finsler-like 
geometries  
\cite{stavrinos-ikeda 1999,stavrinos-ikeda 2000,Stavrinos:2012ty,kour-stath-st 
2012,Stavrinos:2016xyg,Triantafyllopoulos:2018bli,Minas:2019urp} 
(note that the term ``spacetime anisotropy'' in this framework is related to the Lorentz 
violation features of the geometry and should not be 
confused with the spatial anisotropy that can exist   in Riemannian geometry too, e.g.
in Bianchi cases). In specific subclasses of the theory, one can show that these 
novel features are quantified by   effective degrees of freedom that behave  as  scalars 
under coordinate transformations. Hence, one can obtain an effective 
scalar-tensor 
theory arising from a Lorentz fiber bundle. The research of the Finslerian 
scalar-tensor 
theory of gravitation started in 
\cite{stavrinos-ikeda 1999,stavrinos-ikeda 2000} and later in \cite{Stavrinos:2016xyg},
giving an extension of Riemannian ansatz of the scalar-tensor theory of 
gravitation. Additionally, since Finslerian theories exhibit in general 
a violation of Lorentz invariance, one may examine their relation with 
  other classes of  theories that have a preferred vector field or break Lorentz 
symmetry, such as Horava-Lifshitz gravity, Einstein-aether theories   and field 
theories with Lorentz-violating matter fields
\cite{Vacaru:2010rd,Foster:2015yta,Edwards:2018lsn}.

In the present work we are interested in investigating the cosmological applications of 
the  scalar-tensor theory that arises from a Lorentz fiber bundle. As we will see, the 
construction of a Lorentz fiber bundle can provide us an alternative approach of the 
cosmological dynamics of scalar-tensor gravitational theory. In this framework one 
obtains 
extra terms in the Friedmann equations that can lead to interesting cosmological results.

The plan of the work is the following: In Section \ref{GeomBack} we present the
geometrical background of our construction, namely referring to the Lorentz fiber bundle 
structure,
the role of the connection and the geodesics. In Section \ref{Scalartensor} we 
introduce the actions of the theory and we show how a scalar-tensor theory can 
effectively 
appear from the structure of a fiber bundle. Then in Section \ref{Cosmology} we 
investigate in detail the
cosmological applications of the constructed theories, both at late and early times. 
Finally, 
we summarize the obtained results in Section \ref{Conclusions}.

\section{Geometrical Background}
\label{GeomBack}

In this section we briefly review the basic features of Finsler geometry.

\subsection{Basic structure of the Lorentz fiber bundle}

Let us   present the geometrical framework under consideration (a detailed investigation 
can be found in \cite{stavrinos-ikeda 1999,stavrinos-ikeda 2000}).
 We consider a 
6-dimensional Lorentz fiber bundle $E$ 
over a 4-dimensional pseudo-Riemannian spacetime manifold $M$, which locally trivializes 
to 
the form $M \times \{\phi^{(1)}\} \times \{\phi^{(2)}\}$.
The local coordinates on this 
structure are $ (x^\nu,\phi^{(\beta)}) $ with $x^\mu$ the coordinates on the base 
manifold,
where $\kappa, \lambda, \mu,\nu$ take the values from $0$ to $3$, and $\phi^{(\beta)}$ 
the 
coordinates on the fiber where $\alpha,\beta,\gamma$ take the values $1$ and $2$. A 
coordinate transformation on the fiber bundle maps the old coordinates to the new as:
\begin{align}
x^{\nu} \mapsto & x^{\mu'}(x^\nu) \label{x-transormation} \\
\phi^{(\alpha)} \mapsto & \phi^{(\alpha')} = \delta^{(\alpha')}_{(\alpha)}\phi^{(\alpha)},
\end{align}
where $ \delta^{(\alpha')}_{(\alpha)} $ is the Kronecker symbol for the corresponding 
values, and the Jacobian matrix $\pder{x^{\mu'}}{x^\nu}$ is non-degenerate.

Moreover, the space is equipped with a nonlinear connection with local components 
$N^{(\alpha)}_\mu(x^\nu,\phi^{(\beta)})$. The nonlinear connection plays a 
fundamental role in the theory of Lorentz tangent 
bundle and 
vector bundles \cite{Triantafyllopoulos:2018bli,Minas:2019urp,Stavrinos:2012ty}. It is a 
geometrical structure that connects the external-$x$ (horizontal) spacetime with the 
internal-$y$ (vertical) space. This nonlinear connection induces a unique 
split 
of the total space $TE$ into a horizontal distribution $HE$ and a vertical distribution 
$VE$, with
\begin{equation}
TE = HE \oplus VE.
\end{equation}

The adapted basis to this split is
\begin{equation}
\{X_K\} = \{\delta_\mu = \partial_\mu - 
N^{(1)}_\mu\overline\partial_{(1)} - N^{(2)}_\mu\overline\partial_{(2)}, 
\overline\partial_{(\alpha)} \},
\end{equation}
with $\partial_\mu = 
\pder{}{x^\mu}$ and $ \overline\partial_{(\alpha)} = \pder{}{\phi^{(\alpha)}} $. The 
vectors 
$\delta_\mu$ span the horizontal distribution, while $\overline\partial_{(\alpha)}$ span 
the vertical distribution.
Furthermore, the dual basis is
\begin{align}
\{X^M\} = \{\de x^\mu,\, & \delta\phi^{(1)} = \de\phi^{(1)} + N^{(1)}_\nu\de 
x^\nu,\nonumber\\
&\delta\phi^{(2)} = \de\phi^{(2)} + N^{(2)}_\nu\de x^\nu\},
\end{align}
with a summation implied over the 
possible values of $\nu$. Capital indices $K,L,M,N,Z\ldots$ span all the range of values 
of indices on a fiber bundle's tangent space. Finally, the basis vectors transform as:
\begin{equation}
\delta_{\mu'} = \pder{x^\mu}{x^{\mu'}}\delta_\mu
\end{equation}
\begin{equation}
\overline\partial_{(\alpha')} = \delta^{(\alpha)}_{(\alpha')} 
\overline\partial_{(\alpha)},
\end{equation}
where a summation is implied over the possible values of $\alpha$. From these we finally 
acquire the 
transformation of the nonlinear connection components as
\begin{equation}
N^{(\alpha')}_{\mu'} = \pder{x^\mu}{x^{\mu'}}\delta^{(\alpha')}_{(\alpha)} 
N^{(\alpha)}_\mu.
\end{equation}

\subsection{Metric tensor and linear connection}

The metric structure of the space is defined as
\begin{equation}\label{metric}
\mathbf G = g_{\mu\nu}(x)\,\de x^\mu\otimes\de x^\nu + 
v_{(\alpha)(\beta)}(x)\,\delta\phi^{(\alpha)}\otimes \delta\phi^{(\beta)},
\end{equation}
where $g_{\mu\nu}$ has a Lorentzian signature $(-,+,+,+)$. As we can see the fiber 
variables $\phi^{(\alpha)}$ play the role of internal variables. The metric 
components 
for the 
fiber coordinates are set as $v_{(0)(0)} = v_{(1)(1)} = \phi(x^\mu)$ and $v_{(0)(1)} = 
v_{(1)(0)} = 0 $. The inverse metric components are defined by the relations 
$g^{\mu\kappa}g_{\kappa\nu} = \delta^\mu_\nu$ and 
$v^{(\alpha)(\gamma)}v_{(\gamma)(\beta)} 
= \delta^{(\alpha)}_{(\beta)}$. Additionally, the transformation rules for the metric 
are calculated as
\begin{gather}
g_{\mu'\nu'} = \pder{x^\mu}{x^{\mu'}}\pder{x^\nu}{x^{\nu'}} g_{\mu\nu} \lin
v_{(\alpha')(\beta')} = \delta^{(\alpha)}_{(\alpha')}\delta^{(\beta)}_{(\beta')} 
v_{(\alpha)(\beta)}.
\end{gather}
Note that the last relation implies that the metric components on the fiber are scalar 
functions of 
$x^\mu$, since under a coordinate transformation we acquire $v_{(1')(1')} = v_{(1)(1)}$ 
and 
$v_{(2')(2')} = v_{(2)(2)}$, or equivalently $\phi(x^{\mu'})=\phi(x^\mu)$. Hence, as we 
mentioned in the Introduction, the internal structure of the Lorentz fiber bundle
induces scalar degrees of freedom. This feature lies in the center of the analysis of 
this work.

One can define a linear connection $D$   in this space, where the following 
rules 
hold:
\begin{align}
& D_{\delta_\nu}\delta_\mu = L^\kappa_{\mu\nu}\delta_\kappa    & 
&D_{\delta_\nu}\overline\partial_{(\alpha)} = L^{(\gamma)}_{{(\alpha)}\nu} 
\overline\partial_{(\gamma)} \label{Dh} \lin
& D_{\overline\partial_{(\beta)}} \delta_\mu = 
C^{(\gamma)}_{\mu(\beta)}\overline\partial_{(\gamma)}   & 
&D_{\overline\partial_{(\beta)}}\overline\partial_{(\alpha)} = 
C^\kappa_{(\alpha)(\beta)}\delta_\kappa \label{Dv}.
\end{align}
Differentiation of the inner product $D_{X_K}<X^M,X_N> = 0$ and use of 
\eqref{Dh},\eqref{Dv} leads to the rules:
\begin{align}
& D_{\delta_\nu}\de x^\kappa = -L^\kappa_{\mu\nu}\de x^\mu &  
&D_{\delta_\nu}\delta\phi^{(\gamma)} = - L^{(\gamma)}_{{(\alpha)}\nu} 
\delta\phi^{(\alpha)} \lin
& D_{\overline\partial_{(\beta)}} \de x^\kappa = - C^\kappa_{(\alpha)(\beta)}\delta 
\phi^{(\alpha)}\! &  &D_{\overline\partial_{(\beta)}}\delta\phi^{(\gamma)} = - 
C^{(\gamma)}_{\mu(\beta)} \de x^\mu.
\end{align}
It is apparent from the above relations that $D_{\delta_\nu}$ preserves the horizontal 
and 
vertical distributions, while $D_{\overline\partial_{(\beta)}}$ maps one to the other.

Following the above rules, covariant differentiation of a vector $V = V^\mu\delta_\mu + 
V^{(\alpha)}\overline\partial_{(\alpha)}$ along a horizontal direction gives:
\begin{align}
    D_{\delta_\nu}V & = \left(\delta_\nu V^\mu + V^\kappa L^\mu_{\kappa\nu} 
\right)\delta_\mu + \left(\delta_\nu V^{(\alpha)} + 
V^{(\gamma)}L^{(\alpha)}_{(\gamma)\nu}\right)\overline\partial_{(\alpha)} \nonumber\\
    & = D_\nu V^\mu \delta_\mu + D_\nu V^{(\alpha)}\overline\partial_{(\alpha)},
\end{align}
where we have defined
\begin{align}
    & D_\nu V^\mu = \delta_\nu V^\mu + V^\kappa L^\mu_{\kappa\nu} \\
    & D_\nu V^{(\alpha)} = \delta_\nu V^{(\alpha)} + 
V^{(\gamma)}L^{(\alpha)}_{(\gamma)\nu}.
\end{align}
Similarly, for the covariant differentiation of $V$ along a vertical direction we obtain
\begin{align}
D_{\overline\partial_{(\beta)}}V  = & \left[\overline\partial_{(\beta)} V^\mu +  
V^{(\alpha)}C^\mu_{(\alpha)(\beta)} 
\right]\delta_\mu \nonumber\\
&  + \left[\overline\partial_{(\beta)} V^{(\alpha)} + V^\mu 
C^{(\alpha)}_{\mu(\beta)}\right]\overline\partial_{(\alpha)} \nonumber\\
 = &\, D_{(\beta)} V^\mu \delta_\mu + D_{(\beta)} 
V^{(\alpha)}\overline\partial_{(\alpha)},
\end{align}
where we have defined
\begin{align}
    & D_{(\beta)} V^\mu = \overline\partial_{(\beta)} V^\mu +  
V^{(\alpha)}C^\mu_{(\alpha)(\beta)} \\
    & D_{(\beta)} V^{(\alpha)} = \overline\partial_{(\beta)} V^{(\alpha)} + V^\mu 
C^{(\alpha)}_{\mu(\beta)}.
\end{align}
Finally, the covariant derivative for a tensor of general rank is obtained in a similar 
way.

The nonzero components for a metric-compatible connection 
$\left\{\Gamma^K_{LN}\right\} = \left\{L^\kappa_{\mu\nu}, L^{(\gamma)}_{{(\alpha)}\nu}, 
C^{(\gamma)}_{\mu(\beta)}, C^\kappa_{(\alpha)(\beta)}\right\}$,
with $L^\kappa_{[\mu\nu]} 
= 
C^\kappa_{[(\alpha)(\beta)]} = 0$,
are uniquely calculated as:
\begin{align}
& L^\kappa_{\mu\nu} = \gamma^\kappa_{\mu\nu} \label{conn lh} \\
& L^{(\gamma)}_{{(\alpha)}\nu} = \frac{1}{2\phi}\partial_\nu\phi 
\delta^{(\gamma)}_{(\alpha)} \label{conn lv} \\
& C^{(\gamma)}_{\mu(\beta)} = \frac{1}{2\phi}\partial_\mu\phi \delta^{(\gamma)}_{(\beta)} 
\label{conn ch} \\
& C^\kappa_{(\alpha)(\beta)} = -\frac{1}{2}g^{\kappa\nu}\partial_\nu\phi 
\delta_{(\alpha)(\beta)} \label{conn cv},
\end{align}
with 
\begin{equation}
     \gamma^\kappa_{\mu\nu} = \frac{1}{2}g^{\kappa\lambda}\big(\partial_\mu 
g_{\lambda\nu} + \partial_\nu g_{\lambda\mu} - \partial_\lambda g_{\mu\nu}\big),
\end{equation}
while $\delta^{(\gamma)}_{(\alpha)}$, $\delta_{(\alpha)(\beta)}$ etc. are Kronecker 
symbols. 

Now, the curvature tensor of a linear connection is defined as the vector-valued map
\begin{equation}
\tilde R(V,Y)Z = D_V D_Y Z - D_Y D_V Z - D_{[V,Y]}Z,
\end{equation}
where $V,Y,Z$ are vector fields on $E$. Its local components in the adapted basis are 
defined as
\begin{equation}
\tilde R^K_{LMN}X_K = \tilde R(X_M,X_N)X_L,
\end{equation}
where
\begin{align}
\tilde R^K_{LMN} = & X_M\Gamma^K_{LN} - X_N\Gamma^K_{LM} + \Gamma^Z_{LN}\Gamma^K_{ZM}- 
\Gamma^Z_{LM}\Gamma^K_{ZN} \nonumber \\ & + \Gamma^K_{LZ}\mathcal W^Z_{NM},
\end{align}
with $\mathcal W^Z_{NM}X_Z = [X_N,X_M]$.
The generalized Ricci tensor is then defined as
\begin{equation}
\tilde R_{MN} = \tilde R^K_{MKN},
\end{equation}
and the corresponding scalar curvature is
$\tilde R = g^{\mu\nu}\tilde R_{\mu\nu} + v^{(\alpha)(\beta)} \tilde R_{(\alpha)(\beta)}$.
For the linear connection components (\ref{conn lh})-(\ref{conn cv}) one obtains:
\begin{align}
&\tilde R_{\mu\nu} = R_{\mu\nu} + \frac{1}{2\phi^2}\partial_\mu\phi\partial_\nu\phi - 
\frac{1}{\phi}D_\mu D_\nu\phi + \frac{1}{\phi}\partial_\mu\phi 
\overline\partial_{(\alpha)} N^{(\alpha)}_\nu, \\
&\tilde R_{(\alpha)(\beta)} = -\frac{1}{2}\square\phi \delta_{(\alpha)(\beta)},
\end{align}
with $R_{\mu\nu} = \partial_\kappa\gamma^\kappa_{\mu\nu} - 
\partial_\nu\gamma^\kappa_{\mu\kappa} + 
\gamma^\lambda_{\mu\nu}\gamma^\kappa_{\lambda\kappa} - 
\gamma^\lambda_{\mu\kappa}\gamma^\kappa_{\lambda\nu}$ the Ricci tensor of the 
Levi-Civita connection, and $\square = D^\mu D_\mu$. 
Lastly, the scalar curvature $\tilde R=g^{\mu\nu}\tilde R_{\mu\nu} + 
v^{(\alpha)(\beta)}\tilde R_{(\alpha)(\beta)}$ 
 is then
\begin{equation}\label{fiber scalar curvature}
\tilde R = R - \frac{2}{\phi}\square \phi + 
\frac{1}{2\phi^2}\partial_\mu\phi\partial^\mu\phi + 
\frac{1}{\phi}\partial^\mu\phi\overline\partial_{(\alpha)}N^{(\alpha)}_\mu.
\end{equation}
with $R=g^{\mu\nu}R_{\mu\nu}$.

From the above it becomes clear that the internal properties of the Lorentz fiber bundle
eventually induce a scalar-tensor structure. This is the central feature of our work, and 
it will be later investigated in a cosmological framework.

\subsection{Geodesics}

A tangent vector to a curve $\gamma(\tau)$ is written as:
\begin{align}
    Y & = \frac{dx^\mu}{d\tau}\partial_\mu + 
\frac{d\phi^{(\alpha)}}{d\tau}\overline\partial_{(\alpha)}\nonumber  = 
\frac{dx^\mu}{d\tau}\delta_\mu + 
\frac{\delta\phi^{(\alpha)}}{d\tau}\overline\partial_{(\alpha)} \nonumber\lin
      & = Y^\mu\delta_\mu + Y^{(\alpha)}\overline\partial_{(\alpha)},
\end{align}
with $Y^\mu = \dfrac{dx^\mu}{d\tau}$ and $Y^{(\alpha)} = 
\dfrac{\delta\phi^{(\alpha)}}{d\tau} = \dfrac{d\phi^{(\alpha)}}{d\tau} + 
N^{(\alpha)}_\nu\dfrac{dx^\nu}{d\tau}$. Geodesics are the curves with an autoparallel 
tangent vector, namely
\begin{equation}
    D_YY = 0.
\end{equation}
This relation leads to the following geodesic equations:
\begin{gather}
    \frac{d^2x^\mu}{d\tau^2} + \gamma^\mu_{\kappa\lambda} 
\frac{dx^\kappa}{d\tau}\frac{dx^\lambda}{d\tau} = 0, \label{geodesics 1}\lin
    \frac{d\phi^{(\alpha)}}{d\tau} = - N^{(\alpha)}_\mu \frac{dx^\mu}{d\tau} 
\label{geodesics 2}.
\end{gather}
Relation \eqref{geodesics 1} is identical with the geodesics equation of general 
relativity. The new equation of Finsler-like geometry is \eqref{geodesics 2}, which 
relates 
the fiber velocity 
$ \dot\phi^{(\alpha)} $ with the velocity on spacetime $\dot x^\mu $ via the nonlinear 
connection $ N^{(\alpha)}_\mu $.   From geodesic 
equations (\ref{geodesics 1}),(\ref{geodesics 2}) one can see that test 
particles obey the weak equivalence principle. However, Finsler   and 
Finsler-like theories, similarly to general scalar-tensor theories,  may 
violate the strong equivalence principle  
\cite{Puetzfeld:2015jha,Alberte:2019llb,Barausse:2017gip}. Hence, in the 
end one should check whether these violations are inside the corresponding 
experimental bounds.

\section{Scalar tensor theories from the Lorentz fiber bundle}
\label{Scalartensor}

\subsection{Field equations}

Having presented the foundations and the underlying structure of this form of Finsler-like 
geometry, in 
this section we proceed by constructed physical theories. In particular, we can write 
an action as 
\begin{equation}
\mathcal S_G = \frac{1}{16\pi G}\int \sqrt{|\det\mathbf G|}\,\mathcal L_G dx^{(N)},
\end{equation}
where $dx^{(N)} = d^4x\wedge\de\phi^{(1)}\wedge\de\phi^{(2)}$. Following  
\cite{Minas:2019urp} we will  consider two cases of Lagrangian densities:
\begin{enumerate}

\item A Lagrangian density of the form
\begin{equation}\label{lagrangian holonomic}
\mathcal L_G = \mathcal{\tilde R} - \frac{1}{\phi}V(\phi),
\end{equation}
where $V(\phi)$ is a potential for the scalar $\phi$, and
\begin{equation}
    \mathcal{\tilde R} = R - \frac{2}{\phi}\square \phi + 
\frac{1}{2\phi^2}\partial_\mu\phi\partial^\mu\phi
\end{equation}
is the curvature for the specific case of a holonomic basis $ 
[X_M,X_N] = 0 $, in order for the last term in \eqref{fiber scalar curvature} to 
vanish.
Note that the presence of the scalar field in the denominators, that will 
be also transferred to the field equations below, requires to choose the 
potential in a suitable way in order for the dynamics not to lead to 
divergences. This is a requirement that holds for 
the prototype of scalar-tensor theories, namely the Brans-Dicke theory, as well 
as for large classes of Horndeski theories in which scalar-field-related terms 
appear in denominators.

\item A Lagrangian density of the form
\begin{equation}\label{lagrangian nonholonomic}
\overline{\mathcal L}_G = \tilde R
\end{equation}
on a non-holonomic basis, where the noninear connection components are considered as 
functions of $\phi$, $\partial_\mu\phi$ and $g_{\mu\nu}$.
\end{enumerate}

Additionally, and in order to eventually investigate cosmological applications, we   add 
the matter sector too, considering the total action
\begin{equation}
\mathcal S = \frac{1}{16\pi G} \int \sqrt{|\det\mathbf G|}\,\mathcal L_G dx^{(N)}+ \int 
\sqrt{|\det\mathbf G|}\,\mathcal L_M dx^{(N)}.
\label{totaction}
\end{equation}
Since for the determinants  $\det\mathbf G$ and $\det g$ we have the relation 
$\det\mathbf 
G=\phi^2 \det g $, the above total action can be re-written as
\begin{equation}
\mathcal S =  \frac{1}{16\pi G}\int \sqrt{|\det g|}\,\phi\mathcal L_G  dx^{(N)} + \int 
\sqrt{|\det g|}\,\phi\mathcal L_M dx^{(N)}.
\label{totaction2}
\end{equation}
Finally, the same relation holds for the second case, namely for the action of 
$\overline{\mathcal L}_G$.

For the first case we insert the Lagrangian density \eqref{lagrangian holonomic} into the 
action \eqref{totaction2} 
and we vary with respect to $g_{\mu\nu}$ and $\phi$. For $\delta S = 0$ we extract the 
equations:
\begin{align}
    &  E_{\mu\nu} + \frac{1}{\phi}g_{\mu\nu} \left(\square\phi - \frac{1}{4\phi} 
\partial^\lambda\phi\partial_\lambda\phi\right) - \frac{1}{\phi}D_\mu D_\nu\phi 
\nonumber\\
    & + \frac{1}{2\phi^2} \partial_\mu\phi \partial_\nu \phi + \frac{1}{2\phi}g_{\mu\nu}V 
    =  8\pi G T_{\mu\nu} \label{hol feq 1},
\end{align}
where $G$ is the gravitational constant, $E_{\mu\nu} = R_{\mu\nu} - \frac{1}{2}R 
g_{\mu\nu}$
is the Einstein tensor, and $T_{\mu\nu} = 
-\dfrac{2}{\sqrt{|\det g|}}\dfrac{\delta(\sqrt{|\det g|}\mathcal L_M)}{\delta 
g^{\mu\nu}}$ is the energy-momentum tensor. Additionally, the scalar-field equation of 
motion reads as
\begin{align}
    \square\phi = \phi\big(R-V'\big) + 
\frac{1}{2\phi}\partial^\lambda\phi\partial_\lambda\phi 
    + 16\pi G \phi \mathcal L_M \label{hol feq 2},
\end{align}
where a prime denotes differentiation with respect to $\phi$.

For the second case we use the Lagrangian density \eqref{lagrangian nonholonomic}, i.e. 
the 
full scalar curvature \eqref{fiber scalar curvature} without any potential for the 
scalar. 
Variation of the action 
\eqref{totaction2} with respect to $g_{\mu\nu}$ and $\phi$ gives the equations:
\begin{align}
    & E_{\mu\nu} + \frac{1}{\phi}g_{\mu\nu} \left(\square\phi - \frac{1}{4\phi} 
\partial^\lambda\phi\partial_\lambda\phi  - \frac{1}{2}N_\lambda\partial^\lambda\phi 
\right)
    - \frac{1}{\phi}D_\mu D_\nu\phi \nonumber\\
    & + \frac{1}{2\phi^2} \partial_\mu\phi \partial_\nu \phi + 
\frac{1}{\phi}N_{(\mu}\partial_{\nu)}\phi  + \frac{1}{\phi}\pder{N_\lambda}{g^{\mu\nu}}\partial^\lambda\phi 
 = 8\pi G T_{\mu\nu} \label{nonhol feq 1},
\end{align}
with
\begin{equation}\label{N def}
  N_\lambda = \overline\partial_{(\alpha)}N^{(\alpha)}_\lambda,
\end{equation}
 and
\begin{align}
     \square\phi = & \, \phi R + \frac{1}{2\phi}\partial^\lambda\phi\partial_\lambda\phi 
- 
\phi D^\mu N_\mu - \phi \pder{N_\nu}{(\partial_\mu\phi)}D_\mu D^\nu\phi \nonumber\\
     & +   \phi\left[\pder{N_\nu}{\phi} - 
D_\mu\left(\pder{N_\nu}{(\partial_\mu\phi)}\right) \right] \partial^\nu\phi    + 16\pi G \phi \mathcal L_M 
\label{nonhol feq 2}.
\end{align}
Hence, in order to proceed   we need to consider a   specific form of $N_\lambda$. 
We 
choose the following general form:
\begin{equation}\label{Nl}
    N_{\lambda} = \frac{A(\phi)}{2\phi} \partial_\lambda\phi,
\end{equation}
where $A(\phi)$ is a real function of $\phi$.
Taking into account relations \eqref{N def} and \eqref{Nl} we acquire the new form of the 
nonlinear connection components, namely $
    N_\lambda^{(\alpha)} = \frac{A(\phi)}{2\phi} \partial_\lambda\phi\, \phi^{(\alpha)}
$. Thus, substitution into equations \eqref{nonhol feq 1} and \eqref{nonhol feq 2} gives:
\begin{align}
     & E_{\mu\nu} + \frac{1}{\phi}g_{\mu\nu} \left[\square\phi - 
\left(\frac{1+A}{4\phi}\right) 
\partial^\lambda\phi\partial_\lambda\phi\right] - \frac{1}{\phi}D_\mu D_\nu\phi 
\nonumber\\
     & \ \ \ \ \ +\left( \frac{1+A}{2\phi^2}\right) \partial_\mu\phi \partial_\nu 
\phi = 8\pi G T_{\mu\nu} 
\label{nonholfin feq 1},
\end{align}
and
\begin{align}
     (1+A)\square\phi = & \, \phi R + \frac{1}{2\phi}\left(1+A+\phi A' \right) 
\partial^\lambda\phi\partial_\lambda\phi \nonumber\\
     & - \partial^\lambda\phi\partial_\lambda A + 16\pi G \phi \mathcal L_M 
\label{nonholfin feq 2},
\end{align}
where a prime denotes differentiation with respect to $\phi$.

We close this subsection by discussing whether the constructed theory is free 
from 
pathologies.  In general, in any theory first 
of all one needs to examine whether there are Ostrogradsky ghosts, namely 
whether the theory has higher-order (time) derivatives in the general equations 
of motion. Note that if the theory has second-order field equations then it is 
guaranteed that it does not have Ostrogradsky ghosts, however if it does have 
higher-order field equations it is not guaranteed that it does have
Ostrogradsky ghosts,  since the higher-order terms may correspond to extra 
degrees of freedom with second-order field equations (for instance this is the 
case of $f(R)$ gravity as the Hamiltonian analysis reveals). 

As one can see, for the first case of the Lagrangian density \eqref{lagrangian 
holonomic}, namely  in the equations of motion (\ref{hol feq 1}),(\ref{hol feq 
2}), higher than second-order derivatives are absent. Similarly, for the second 
case of the Lagrangian density \eqref{lagrangian nonholonomic}, namely in  the 
equations of motion (\ref{nonhol feq 1}),(\ref{nonhol feq 2}),  imposing the 
 non-linear connection (\ref{Nl}) (which does not 
include higher than first derivatives) leads to field equations 
\eqref{nonholfin feq 1},\eqref{nonholfin feq 2} where higher than second-order 
derivatives are absent.   Hence, we deduce that our theory satisfies the basic 
requirement to be free from Ostrogradsky ghosts.

Nevertheless, it is well known that even if a theory is     free from 
Ostrogradsky ghosts, still 
other kinds of pathologies, like ghost and Laplacian instabilities, may arise, 
related to the perturbation analysis,    around a general or particular 
background(s). In order to examine whether the theory at hand exhibits such 
instabilities a full perturbation 
analysis is needed.  However, the present work is devoted to a first 
study of the cosmological 
applications of the theory, and hence we focus on the basic requirement that 
the theory is free from Ostrogradsky ghosts. The full investigation of  ghosts 
and Laplacian instabilities lies beyond the scope of the work and it is left 
for a future project.

\subsection{Energy-momentum conservation}
 
	In this subsection we investigate the energy-momentum conservation in our 
model. A full analysis goes beyond the scope of this manuscript, however with 
some simple reasoning we can extract important results.

In Einstein's general relativity, 
diffeomorphism invariance of the theory leads to energy-momentum conservation 
with respect to the Levi-Civita connection, through the Bianchi identities.
	In order to examine what happens in our case we may consider 
infinitesimal local diffeomorphisms on the Lorentz fiber-bundle induced by a 
vector field:
	\begin{equation}
	\xi = \xi^\mu(x)\, \delta_\mu + \xi^{(\alpha)} \overline\partial_{(\alpha)},
	\end{equation}
	where the spacetime components of the $\xi$ vector are only $x-$dependent  
in order to be compatible with the transformation group of $x-$coordinates given 
in  \eqref{x-transormation}.
Thus, a point with spacetime coordinates $x^\mu$ is mapped via the 
diffeomorphism to a point with spacetime coordinates $x'^\mu = x^\mu + \epsilon 
\xi^\mu(x)$, where $\epsilon$ is sufficiently small. 

We consider the metric  \eqref{metric}. Since  $g_{\mu\nu}(x)$ 
depends only on $x^\mu$ and not on $\phi^{(\alpha)}$, mathematically it 
behaves like a pseudo-Riemannian metric of spacetime. The 
transformation rule for this kind of metric under a diffeomorphism is
	\begin{equation}\label{g transformation}
	g'_{\mu\nu} = g_{\mu\nu} - 2\epsilon \nabla_{(\mu} \xi_{\nu)},
	\end{equation}
	where $\nabla$ is the Levi-Civita connection and the parenthesis denotes  
symmetrization of indices.
	Additionally,  the scalar field $\phi(x)$ transforms as
	\begin{equation}
	\phi' = \phi -\epsilon \xi^\mu \nabla_\mu\phi.
	\end{equation}
	In the following, we will assume that $\phi(x)>0$.
	
	Now, it is   known   that the action
	\begin{equation}
	S_i = \int_V d^n \mathcal U \sqrt{|\det\mathbf G |}\, Q(\mathcal U)
	\end{equation}
	on a closed subspace of some $n-$dimensional  differential manifold with 
local coordinates $\mathcal U^A$, $A=1,\ldots,n$, for a scalar function 
$Q(\mathcal U)$, is invariant under a local diffeomorphism $\xi$ that vanishes 
at the boundary $\partial V$. Hence, applying the above local diffeomorphism on 
the variation of the matter action part of    \eqref{totaction}, namely on 
$\delta S_M  =  \delta \int_V dx^{(N)} \phi\sqrt{|\det g|} \mathcal L_M = 
0$, we obtain:
	\begin{align}
	   & \int_V dx^{(N)}\phi \sqrt{|\det g|}\, 
T^{\mu\nu}\nabla_\mu\xi_\nu \nonumber\\
	& - \int_V dx^{(N)}\sqrt{|\det g|} \mathcal L_M \xi^\mu \partial_\mu \phi \nonumber \\
	= & \int_{\partial V} dx^{(N-1)}\sqrt{|\det\gamma|}\, T^{\mu\nu}\xi_\nu n_\mu \nonumber\\
	& - \int_V dx^{(N)}\sqrt{|\det g|}\, \phi \, \xi_\nu \nabla_\mu T^{\mu\nu} \nonumber\\
	& - \int_V dx^{(N)}\sqrt{|\det g|}\, \phi \, \frac{\mathcal L_M}{\phi} 
\xi^\mu \partial_\mu \phi = 0  \label{matter fields transformation},
	\end{align}
	where we have used the definition of the matter energy-momentum tensor  
$T_{\mu\nu} = 
-\dfrac{2}{\sqrt{|\det g|}}\dfrac{\delta(\sqrt{|\det g|}\mathcal L_M)}{\delta 
g^{\mu\nu}}$, and 
	where $\det\gamma$ is the determinant of the induced metric and $n_\mu$ a 
normal vector field at the boundary $\partial V$. Setting that $\xi^\mu = 0$ at 
$\partial V$,   the first term after the last equality in 
 \eqref{matter fields transformation} vanishes. Additionally, since  
$\xi^\nu$ can be chosen arbitrarily, the last relation finally implies:
	\begin{equation}
	\nabla_\mu T^{\mu}_{\,\,\,\nu} = -\frac{\partial_\nu 
		\phi}{\phi} \mathcal L_M .
\label{Tmunuinteraction}
	\end{equation}
	This expression denotes a departure from general relativity.  Specifically, 
the nonminimal coupling of $\phi$ to the matter fields in the action \eqref{totaction2} induces an 
interaction term in the above generalization of energy-momentum conservation. 
This interaction will have interesting implications in the cosmological 
application of the next section. Finally, we mention that for a constant 
scalar field $\phi$ this relation reduces to the standard conservation law of 
general relativity.

\section{Cosmology}
\label{Cosmology}

In the previous section we constructed the physical theories, namely the actions, on the 
framework of Finsler geometry, considering the cases of holonomic (Lagrangian 
(\ref{lagrangian holonomic})) and non-holonomic (Lagrangian 
(\ref{lagrangian nonholonomic})) basis separately, and we extracted the field equations. 
In 
order to apply them in a cosmological framework we 
 consider a homogeneous and isotropic spacetime with the bundle metric
\begin{align}
    \mathbf{G} = & \, -\de t\otimes\de t  + a^2(t)\left(\de x\otimes\de x + \de 
y\otimes\de y + \de z\otimes\de z\right) \nonumber \\
    & \, + \phi(t)\left(\delta\phi^{(1)}\otimes\delta\phi^{(1)} + 
\delta\phi^{(2)}\otimes\delta\phi^{(2)}\right) \label{bundle metric}.
\end{align}
The first line of \eqref{bundle metric} is the standard spatially flat  
Friedmann-Robertson-Walker (FRW)  metric,
while 
the 
second line arises from the additional structure of the Lorentz fiber bundle.
Moreover, we consider the energy-momentum tensor of the matter  perfect fluid:
\begin{equation} \label{perf fluid}
    T_{\mu\nu} = (\rho_m+P_m)u_\mu u_\nu + P_mg_{\mu\nu},
\end{equation}
with $\rho_m$ the energy density, $P_m$ the pressure and $u^\mu$ the bulk 
4-velocity of the fluid.
As usual, the 
first line of \eqref{bundle metric} defines the comoving frame on this spacetime and 
$u^\mu$ is at rest with respect to it. Finally, note that relation \eqref{perf fluid} 
can be derived from a Lagrangian density $\mathcal L_M=\rho_m$, and thus we will use this 
when the matter Lagrangian appears in the field equations.

\subsection{First case: holonomic basis}

For the spacetime \eqref{bundle metric}, and with the perfect fluid \eqref{perf fluid}, 
the field 
equations \eqref{hol feq 1} and \eqref{hol feq 2} of the holonomic case give:
\begin{align}
    & 3H^2 = 8\pi G\rho_m - \frac{\dot\phi^2}{4\phi^2} +\frac{1}{\phi}\left( \frac{V}{2} 
- 
3H \dot\phi\right)  \label{hol fried 1}, \\
    & \dot H = -4\pi G(\rho_m+P_m)  + \frac{\dot\phi^2}{4\phi^2} + \frac{1}{2\phi}\left( 
H\dot\phi- \ddot\phi\right)\label{hol fried 2}, \end{align}
\begin{align}
    & \ddot\phi + 3H\dot\phi = - 16\pi G \phi \mathcal \rho_m - 6\phi\left(\dot H +2H^2 
\right) + \phi V' + \frac{\dot\phi^2}{2\phi} \label{hol fried 3},
\end{align}
where $H=\dot a/a$ is the Hubble parameter and a dot denotes differentiation with respect 
to coordinate time $t$. In the following we investigate these equations in late and early 
times separately.

\subsubsection{Late-time cosmology}

Observing the forms of the   Friedmann equations (\ref{hol fried 1}),(\ref{hol fried 
2}), we deduce that we can write them in the usual form 
\begin{align}
    & 3H^2 = 8\pi G\left(\rho_m +\rho_{DE}\right)   \label{Fr1b} \\
    & \dot H = -4\pi G\left(\rho_m+P_m +\rho_{DE}+P_{DE}
    \right),\label{Fr12b} 
\end{align}
 defining the effective dark energy density and pressure respectively as
\begin{eqnarray}
&&\rho_{DE}
\equiv\frac{1}{8\pi G}\left[
 - \frac{\dot\phi^2}{4\phi^2} +\frac{1}{\phi}\left( \frac{V}{2} 
- 
3H \dot\phi\right) 
\right]
\\
&&p_{DE}\equiv 
-\frac{1}{32\pi G}\left[\frac{2\phi V-8H 
\phi\dot{\phi}+\dot{\phi}^2-4\phi\ddot{\phi}}{\phi^2}\right],
\end{eqnarray}
and thus the corresponding equation-of-state parameter will be
\begin{eqnarray}
w_{DE}\equiv  \frac{P_{DE}}{\rho_{DE}}.
\label{wDEdef}
\end{eqnarray}
Hence, as we mentioned in the introduction, in the present scenario of holonomic 
basis, we obtain an effective dark energy sector that arises from the intrinsic 
properties of the Finsler-like geometry, and in particular from the scalar-tensor theory 
of 
the Lorentz fiber bundle. 

Inserting the definitions of $\rho_{DE}$ and $P_{DE}$ into the scalar field equation of 
motion (\ref{hol fried 3}), and using the Friedmann equations (\ref{hol fried 
1}),(\ref{hol 
fried 2}), we find that 
\begin{eqnarray}
&&\dot{\rho}_m+3H(\rho_m+P_m)=-\frac{2\dot{\phi}}{\phi}\rho_m,
\label{intereq1}\\
&&\dot{\rho}_{DE}+3H(\rho_{DE}+P_{DE})=\frac{2\dot{\phi}}{\phi}\rho_m.
\label{intereq2}
\end{eqnarray}
Interestingly enough,  we find that the scenario at hand induces an 
interaction between the 
matter and dark energy sector, again as a result of the intrinsic geometrical structure 
(this was already clear by the presence of $\rho_m$ in the right-hand-side of (\ref{hol 
fried 3})), with the total energy density being conserved as expected from the 
conservation of the total energy-momentum tensor. Actually, equations 
(\ref{intereq1}),(\ref{intereq2}), derived in a specific framework, reflect the 
general expression 
(\ref{Tmunuinteraction}).
Hence, although the geometrical scalar-tensor terms that appear in the 
Friedmann equations (\ref{hol fried 1}),(\ref{hol 
fried 2}) may be obtained from other theories too, such as the Horndeski theory 
\cite{Horndeski:1974wa} or the theory of generalized galileons 
\cite{Nicolis:2008in,Deffayet:2009mn}, the interaction term is something that does not 
fundamentally exist in these theories. In summary, the scenario at hand does not fall 
into the Horndeski class, and thus it is interesting to examine its cosmological 
implications. 

We mention here that in literature of interacting cosmology, in
general the interaction terms  are imposed by hand, namely one breaks by hand the total 
conservation law
$\dot{\rho}_m+3H(\rho_m+P_m)+\dot{\rho}_{DE}+3H(\rho_{DE}+P_{DE})=0$ into 
$\dot{\rho}_m+3H(\rho_m+P_m)=Q$ and $\dot{\rho}_{DE}+3H(\rho_{DE}+P_{DE})=-Q$, with $Q$ 
the phenomenological descriptor of the interaction that is then chosen at 
will or under specific theoretical justifications
\cite{Barrow:2006hia,Amendola:2006dg,Chen:2008ft,Gavela:2009cy,Chen:2011cy,Yang:2014gza,
Faraoni:2014vra}. However, in the present scenario of scalar-tensor theory from the fiber 
bundle 
with a holonomic basis, we see that such an interaction term, and in particular 
$Q=-\frac{2\dot{\phi}}{\phi}\rho_m$, arises naturally from the Finsler-like structure of 
the 
geometry
(note that although interaction functions of the form $Q \propto \rho_m$ 
have been shown to present instabilities at early times if the involved 
dark-energy sector has a constant equation-of-state parameter 
\cite{Valiviita:2008iv}, in the case of general dark energy equation of state 
such instabilities are absent).
Such a property opens new directions to look at the appearance of the 
interaction terms, especially under the light of its significance to alleviate the 
$H_0$ 
tension \cite{Yang:2018euj,Pan:2019gop},  and it is one of the main results of the 
present 
work.

Let us now investigate the cosmological behavior that is induced from the scenario at 
hand. We elaborate the Friedmann equations (\ref{hol fried 1}),(\ref{hol fried 
2}) numerically,
using as usual  as   independent variable
the
redshift $z$, defined  through  $ 1+z=1/a$ (we set the present scale factor to
$a_0=1$). We use various potential forms, such as linear, quadratic, quartic, and 
exponential.
Moreover, we choose the initial conditions in order to obtain
$\Omega_{DE}(z=0)\equiv\Omega_{DE0}\approx0.69$  and 
$\Omega_m(z=0)\equiv\Omega_{m0}\approx0.31$  in agreement with observations
\cite{Ade:2015xua}.
Furthermore, for  the matter sector we choose $w_m\equiv P_m/\rho_m=0$, namely the 
standard pressureless dust matter.

\begin{figure}[ht]
\includegraphics[scale=0.45]{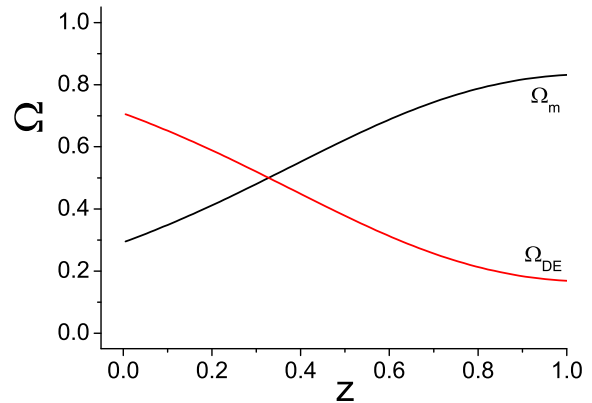}
\caption{
{\it{ The evolution of the
 matter density
parameter $\Omega_{m}$ (black-solid) and of the
effective dark energy     
density parameter $\Omega_{DE}$   (red-solid), as a function of 
the redshift $z$,
in the case of holonomic basis, with quadratic potential $V(\phi)=\alpha 
\phi^2$, for  
$\alpha=0.5$ 
in units where $8\pi G=1$. We have imposed the initial conditions
$\Omega_{DE}(z=0)\equiv\Omega_{DE0}\approx0.69$  in agreement with observations.
}} }
\label{holoOmegas}
\end{figure}
 \begin{figure}[ht]
\includegraphics[scale=0.45]{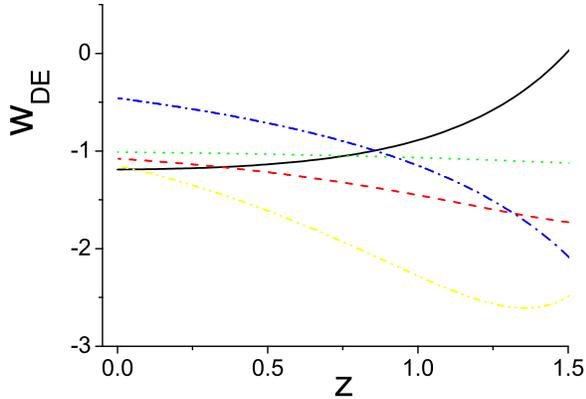}
\caption{
{\it{  The evolution of the equation-of-state parameter  $w_{DE}$  of the 
effective dark
energy of holonomic case, for various choices of the potential $V(\phi)$: 
$V(\phi)=\alpha \phi$ with $\alpha=0.2$ (black - solid), $V(\phi)=\alpha \phi^2$ 
with 
$\alpha=0.5$ 
(red - dashed), $V(\phi)=\alpha \phi^2$ with $\alpha=1$ 
(green - dotted), $V(\phi)=\alpha \phi^4$ with $\alpha=0.1$ 
(blue  - dashed-dotted), $V(\phi)=V_0 e^{\lambda \phi}$ with 
$V_0=1$,$\lambda=2$ 
(yellow - dashed-dotted-dotted).
We have imposed the initial conditions
$\Omega_{DE}(z=0)\equiv\Omega_{DE0}\approx0.69$ and we use units where $8\pi 
G=1$.
}} }
\label{holowde}
\end{figure}

In  Fig. \ref{holoOmegas} we present  $\Omega_{DE}(z)$ and $\Omega_{m}(z)$ for the 
case of quadratic potential $V(\phi)=\alpha \phi^2$, with  
$\alpha=0.5$. As we can see, we can obtain the usual
thermal history of the universe, i.e. the sequence of matter and dark energy epochs,
in agreement with observations, although we have not considered an explicit cosmological 
constant.
Furthermore, in order   to examine the effect of the potential forms on  $w_{DE}$, in  
Fig. \ref{holowde} we
depict the evolution of  $w_{DE}(z)$ for various potential choices, namely for linear 
$V(\phi)=\alpha \phi$, quadratic $V(\phi)=\alpha \phi^2$, quartic $V(\phi)=\alpha \phi^4$ 
and 
exponential $V(\phi)=V_0 e^{\lambda \phi}$. We observe a rich behavior in the evolution 
of $w_{DE}(z)$, which may lie in the quintessence or phantom regime, or experience the 
phantom-divide crossing. These properties are not easily obtained in theories with scalar 
degrees of freedom,  and reveal the capabilities of the scenario.

 \subsubsection{Inflation}
 
We close this subsection by discussing the application of the scenario at hand in the 
case of early times, namely examining the inflationary realization. In this case the 
matter sector can be neglected.  As one can easily see, the equations accept the de 
Sitter solution $a(t)=a_{init} e^{H_0 t} $, which is the basis of any inflationary 
scenario. 
In particular, choosing the linear potential $V=\alpha\phi$ then (\ref{hol fried 
1}),(\ref{hol 
fried 2}),(\ref{hol fried 3}) accept the solution  
 \begin{eqnarray}
&& H(t)=H_0\nonumber\\
&& \phi(t)=\phi_0 e^{2 H_0 t},
\end{eqnarray}
with $\phi_0$ an integration constant, if we choose $\alpha=20H_0^2$. Note that  the 
above de Sitter solution is obtained without considering an explicit 
cosmological constant.

Going into a more 
realistic inflationary realization, with a successful exit after a desired e-folding, and 
the desired scalar spectral index and tensor-to-scalar ratio, we can follow the  method 
applied in \cite{Minas:2019urp} and reconstruct the potential $V(\phi)$ that produces any 
given form of $H(t)$. In particular, imposing the desired inflationary  $H(t)$,  Eq. 
(\ref{hol fried 2}) 
becomes a simple differential equation whose solution provides $\phi(t)$. Inserting this 
$\phi(t)$ into (\ref{hol fried 1}) gives   $V(t)$ as
\begin{eqnarray}
\label{holoinflVsol}
V(t)=6 H(t)[\dot{\phi}(t)+\phi(t)H(t)]+\frac{\dot{\phi}(t)^2}{2\phi(t)}.
\end{eqnarray}
Note that in the absence of matter, Eq. (\ref{hol fried 3}) is not independent from 
(\ref{hol fried 1}),(\ref{hol 
fried 2})  and thus solution of the latter two ensures that  (\ref{hol fried 3}) is 
satisfied too. Hence,   knowing both $\phi(t)$ and $V(t)$ we can eliminate time and 
reconstruct explicitly the  potential $V(\phi)$. Therefore, this potential is the one 
that 
 generates  the initially given inflationary $H(t)$. The capability  of accepting 
inflationary solutions is an additional advantage of the scenario of Lorentz fiber-bundle 
induced 
scalar-tensor theory with holonomic basis.

We close this subsection by mentioning that in the above analysis we have 
indeed described the inflation epoch and the subsequent thermal history of the 
universe with the same model  (relations (\ref{hol fried 1})-(\ref{hol fried 
3})). However,  we have used   different potential and 
parameter 
choices for each case, which is standard in the literature since the  
involved energy scales in principle differ by many orders of magnitude. If one 
desires to construct a scenario that can describe the universe evolution from 
inflation to today in a unified way,   he  must introduce additional 
mechanisms/couplings that may successfully lead to quintessential inflation
\cite{Hossain:2014zma,Geng:2017mic}. It would be interesting to see whether 
both pictures could be   unified  after an extension of the model 
with extra 
mechanisms.

\subsection{Second case: Nonholonomic basis}

For the spacetime \eqref{bundle metric}, and with the perfect fluid \eqref{perf fluid}, 
the field 
equations \eqref{nonholfin feq 1} and \eqref{nonholfin feq 2} give:

\begin{eqnarray}
  &&
  \!\!\!\!  \!\!\!\!\!\!\!\!
  3H^2 = 8\pi G\rho_m - (1+A)\frac{\dot\phi^2}{4\phi^2} - 3H 
\frac{\dot\phi}{\phi} 
\label{nonhol fried 1},\\
&&\!\!\!\!  \!\!\!\!\!\!\!\!
    \dot H = \! -4\pi G(\rho_m\!+\!P_m) + (1\!+\!A)\frac{\dot\phi^2}{4\phi^2} + 
\frac{1}{2\phi} 
\left(H\dot\phi\! - \!\ddot\phi\right)\!, \label{nonhol fried 2}
\\
  && \!\!\!\! \!\! \!\! \!\!\!\!(1+A)\left( \ddot\phi + 3H\dot\phi \right) =  - 16\pi G 
\phi \mathcal \rho_m - 
6\phi\left(\dot H +2H^2 \right) \nonumber\\
    &&\,
    \ \ \     \ \ \     \ \ \     \ \ \     \ \ \     \ \ \     \ \ \     \  \,
    + \frac{\dot\phi^2}{2\phi}\left( 1+A+\phi A' \right) - \dot\phi \dot A
     \label{nonhol fried 3},
\end{eqnarray}
where a dot denotes a derivative with respect to $t$ and a prime denotes a derivative 
with 
respect to $\phi$.

\subsubsection{Late-time cosmology}

Similarly to the holonomic case of the previous subsection, we can re-write the 
Friedmann equations  (\ref{nonhol fried 1}),(\ref{nonhol fried 2}) in the standard form 
 (\ref{Fr1b}),(\ref{Fr12b}), introducing an effective dark energy sector with energy 
density and pressure respectively as  
\begin{eqnarray}
&&\rho_{DE}
\equiv\frac{1}{8\pi G}\left[ - (1+A)\frac{\dot\phi^2}{4\phi^2} - 3H \frac{\dot\phi}{\phi} 
\right]
\\
&&p_{DE}\equiv 
-\frac{1}{32\pi G}\left[\frac{
(1+A)\dot\phi^2-4\phi\ddot{\phi}
-8H 
\phi\dot{\phi}}{\phi^2}\right],
\end{eqnarray}
and thus the corresponding equation-of-state parameter is
$
w_{DE}\equiv   P_{DE}/\rho_{DE}$.
Thus,  in the present scenario of nonholonomic 
basis, we also obtain an effective dark energy sector that arises from the scalar-tensor 
theory of the Lorentz fiber bundle. 

Inserting the above $\rho_{DE}$ and $P_{DE}$ into the scalar field equation of 
motion (\ref{nonhol fried 3}), and using the Friedmann equations (\ref{nonhol fried 
1}),(\ref{nonhol fried 2}), we find that 
\begin{eqnarray}
&&\dot{\rho}_m+3H(\rho_m+P_m)=-\frac{2\dot{\phi}}{\phi}\rho_m,\\
&&\dot{\rho}_{DE}+3H(\rho_{DE}+P_{DE})=\frac{2\dot{\phi}}{\phi}\rho_m.
\end{eqnarray}
Similarly to the holonomic case, in the present non-holonomic scenario the intrinsic 
geometrical structure induces an interaction between the 
matter and dark energy sector, while the total energy density is conserved. Therefore, 
the scenario at hand cannot be naturally obtained from  Horndeski or generalized  
galileons theories. 

We mention that although the above interaction 
term coincides with that of the holonomic case, namely 
$Q=-\frac{2\dot{\phi}}{\phi}\rho_m$, and despite the similarities of the Friedmann 
equations of the two cases, there is not a relation between $V(\phi)$ of the 
holonomic case and $A(\phi)$ of the non-holonomic case that could transform one case to 
the other. Hence, the two scenarios correspond to distinct classes of modified theories.

In order to investigate the cosmological evolution in this scenario one could perform a 
full numerical elaboration similar to the previous subsection, resulting to figures 
similar to Figs. \ref{holoOmegas}   
and  \ref{holowde}. However, one can also extract approximate analytical 
solutions that hold at late times, namely in the dark energy epoch. In particular, as it 
is well 
known, a very general and important solution for the scale factor evolution is the
power-law one, namely 
\begin{eqnarray}
a(t)\propto t^n,
\end{eqnarray}
in which case the Hubble function becomes 
 \begin{eqnarray}
H(t)=\frac{n}{t}.
\end{eqnarray}
As an example we choose a linear form for $A(\phi)$, i.e.   $A(\phi)=\alpha\phi+\beta$. 
In this case one can extract the approximate solutions of Eqs. (\ref{nonhol fried 
1}),(\ref{nonhol fried 2}) and  (\ref{nonhol fried 3})  as
\begin{eqnarray}
\phi(t)&\approx &\phi_0 t^{-3n/2}(t+c_1)\\
\rho_m(t)&\approx& \frac{\rho_{m0}}{(t+c_1)^2}.
\end{eqnarray}
These lead to 
  \begin{eqnarray}
&&
\!\!\!\!\!\!\!\!\!\!\!\!\!
\Omega_{DE}(t)\approx  1-\frac{8\pi G \rho_{m0}\, t^2 }{3n^2(t+c_1)^2}
\\
&&\!\!\!\!\!\!\!\!\!\!\!\!\!
w_{DE}(t)\approx 
\left\{(3n-2)[3 n ( \beta-7) - 2 ( \beta+1)]t
\right\}^{-1}\nonumber\\
&&
\ \ \ \ \ \ 
\cdot
\left\{
-2 c_1 n [22 + 6 \beta - 3n(3\beta+6)]\right.
\nonumber\\
&&\ \ \ \ \ \ \ \ \,
\left.+
( 3 n-2)[ n(3\beta+7)-2 (\beta+1) ] t
\right\},
\end{eqnarray}
and thus the asymptotic value of $w_{DE}(t)$  is
  \begin{eqnarray}
 \lim_{t\rightarrow\infty }w_{DE}(t)=
 \frac{ 2(\beta+1) -n(3\beta+7)}
 { 2(\beta+1)-3n (\beta-7) }.
\end{eqnarray}
Hence, we can easily see that the universe at late times exhibits the correct thermal 
history, with the sequence of matter and dark energy epochs and the onset of late-time 
acceleration. Additionally, the dark-energy equation-of-state parameter may be
quintessence-like, phantom-like, or experience the phantom-divide 
crossing during the evolution. Finally, $w_{DE}$ acquires an 
asymptotic value which can lie in the quintessence regime, in the phantom regime, or 
being exactly  equal to the cosmological constant value $-1$ (by choosing 
$\beta=(7n+2)/(3n-2)$). We stress that the above behavior is obtained without considering 
an explicit cosmological constant, namely it arises solely form the fiber bundle
structure of Finsler-like geometry, and this is an additional advantage that reveals the 
capabilities of the scenario.

Let us mention here that in the above example there appears a non-minimal 
coupling between the matter sector  and the scalar field, and such couplings, 
may in general lead to violation of the equivalence principle. Hence, similarly 
to the theories where non-minimal matter couplings are used, such as in 
theories where the matter Lagrangian is coupled to $f(R)$
\cite{Bertolami:2007gv}, in 
$f(R,T)$ gravity with $T$ the trace of the energy momentum tensor 
\cite{Harko:2011kv}, etc, one should choose the involved model parameters in a 
way that the experimental bounds on the equivalence principle are 
satisfied.

 \subsubsection{Inflation}
 
We close this subsection by discussing the application of the scenario at hand in  
inflationary realization. Neglecting the matter sector, we focus on the existence of the 
  de 
Sitter solution $a(t)=a_{init} e^{H_0 t} $. Indeed,    choosing  
$A(\phi)=\alpha$ then (\ref{nonhol fried 1}),(\ref{nonhol 
fried 2}),(\ref{nonhol fried 3}) accept the solution  
 \begin{eqnarray}
&& H(t)=H_0\nonumber\\
&& \phi(t)=\phi_0 e^{-\lambda  t},
\end{eqnarray}
with $\phi_0$ an integration constant, if we choose $\lambda=-3 H_0$, $\alpha=5/3$  or  
$\lambda=-2 H_0$, $\alpha=2$, in units where $8\pi G=1$. Similarly to the holonomic case, 
we mention  
that the above de Sitter solution is obtained without considering an explicit 
cosmological constant.

Finally, we close this analysis by describing the possibility to obtain any  desired
    inflationary  $H(t)$ by suitably choosing the function $A(\phi)$. In particular, 
eliminating $ A$ from 
(\ref{nonhol fried 1}),(\ref{nonhol 
fried 2})
gives the simple differential equation 
\begin{eqnarray}
\label{Fr1nonholdiffbb}
 \ddot{\phi}(t)+5 H(t)\dot{\phi(t)}  +2\phi(t)  
     [ \dot{H}(t)+3 H(t)^2]=0,
\end{eqnarray}
which under the imposed $H(t)$  provides the solution for $\phi(t)$. This $\phi(t)$ 
can be substituted back in   
(\ref{nonhol fried 1}) and provide $A(t)$ as
\begin{eqnarray}
\label{Fr1nonholVt}
A(t)=-12 H(t)\frac{\phi(t)}{\dot{\phi}(t)}\left[
H(t)\frac{\phi(t)}{\dot{\phi}(t)}+1
\right]-1.
\end{eqnarray}
Thus, knowing both $\phi(t)$ and $A(t)$ we can reconstruct the $A(\phi)$ form that 
produces the initially given inflationary $H(t)$.

We close the section by making the following comment. Strictly speaking, a 
more theoretically robust approach to the cosmological investigation would be to
first   construct a potential or a non-linear connection  according to  
theoretical arguments, and then try to examine the induced dynamics. However, 
since the fundamental theory is unknown, in the analysis of 
the present section we followed the widely-used method of reconstructing the 
unknown functions of the theory in order to be consistent with the 
phenomenological requirements. The fact that the resulting forms for the 
potential and the non-linear connection may look complicated is not a problem, 
since this is known to be the case for all viable models of modified gravity 
(e.g. the viable $f(R)$ and $f(T)$ functions in the corresponding theories), 
  under the   requirement to incorporate phenomenology and observational data   
 correctly.

\section{Conclusions}
\label{Conclusions}

In the present work we investigated the cosmological applications of scalar-tensor 
theories that arise effectively from the Lorentz fiber bundle of a Finsler-like geometry. 
The latter is a natural extension of Riemannian one in the case where  the physical 
quantities may  depend on a set of internal variables, too. Hence, in a general 
application of such a construction to a cosmological framework, one obtains extra terms 
in 
the Friedmann equations that can lead to interesting phenomenology. 

We started by a Lorentz fiber bundle structure, where $M \times \{\phi^{(1)}\} \times 
\{\phi^{(2)}\}$ represents a pseudo-Remannian spacetime with two fibers $ \phi^{(1)} $ 
and 
$\phi^{(2)}$. The nonlinear connection under consideration induces a new degree of 
freedom 
that behaves as a scalar under coordinate transformations. Hence, the rich structure of 
Finsler-like geometry can induce an effective scalar-tensor 
theory from the Lorentz fiber bundle.

In the case where a holonomic basis is used, the effective scalar-tensor theory leads 
to the appearance of an effective dark energy sector of geometrical origin
in the Friedmann equations. However, the interesting novel 
feature is that we acquired an interaction between the matter and dark energy sectors, 
arising purely from the internal structure of the theory and not imposed by hand. Hence, 
the theory under consideration cannot be naturally obtained from  Horndeski or 
generalized galileons 
theories. Applying it at late times we found that we can obtain the 
thermal history of the universe, namely the sequence of matter and dark-energy 
epochs, in 
agreement with observations. Additionally, we showed that the effective dark-energy 
equation-of-state parameter can be quintessence-like, phantom-like, or experience the 
phantom-divide crossing during cosmological evolution. These features were obtained 
although we had not considered an explicit cosmological constant, namely they arise 
purely from the intrinsic geometrical structure of Finsler-like geometry, which is a 
significant advantage. Finally, applying the scenario at early times we showed that one 
can acquire an exponential de Sitter solution, as well as obtain an inflationary 
realization with the desired scale-factor evolutions, and thus with the desired 
inflationary observables such as the spectral-index and the tensor-to-scalar ratio.

In the case of a non-holonomic basis we also obtained an effective dark energy sector, 
which moreover exhibits an explicit interaction with the matter sector. Concerning late 
times, we extracted approximate analytical solutions in which the scale factor has a 
power-law evolution. These solutions show the sequence of matter and dark energy epochs, 
in agreement with observations, and furthermore the corresponding dark-energy 
equation-of-state parameter can lie in the quintessence or phantom regime, or experience 
the phantom-divide crossing during the evolution, or even  obtain asymptotically  exactly 
the cosmological constant value. Finally, at early times the scenario at 
hand can also accept de Sitter solutions, as well as a successful inflationary 
realization with the desired  spectral-index and the tensor-to-scalar ratio.

We would like to mention here that the fact that in the present scenario we 
obtain an interacting behavior, as well as a dark energy sector that can lie in 
the phantom regime, may be useful towards the solution of the $H_0$ tension, 
since as it has been investigated in the literature both features may 
successfully lead to its alleviation \cite{Yang:2018euj,Pan:2019gop}.

In summary, the rich structure of Finsler-like geometry can lead to interesting 
cosmological 
phenomenology at both early and late times. 
There are many interesting investigations 
that should be performed along these lines. One should use observational 
data from  Type Ia Supernovae (SNIa), Baryon Acoustic Oscillations (BAO), Cosmic 
Microwave 
Background (CMB) shift parameter and temperature and polarization, direct 
Hubble constant observations, and $f\sigma 8$ data, in order to 
extract constraints on the involved forms and parameters.
In particular, the confrontation with the CMB   power spectrum might be 
quite interesting in   light of the   low-$\ell$ 
anomalies, since  as we mentioned in the introduction Finsler geometry in 
general presents  an intrinsic anisotropy (in the specific construction of 
the present work, which belongs to the more general class of Finsler-like 
geometries,  the intrinsic anisotropy is replaced by the set of internal 
variables $\{\phi^{(1)},\phi^{(2)}\}$).  Nevertheless,  such 
a full confrontation with observations lies beyond the scope
of this work  and it is left for future projects.
Additionally, one could 
perform 
a detailed dynamical analysis in order to reveal the global behavior of the theory with 
Finsler-like cosmology, independently from the initial 
conditions. 
Finally, going beyond the cosmological framework, one could look for  black hole 
solutions in these theories. These necessary studies   are left for future 
investigations.

\section*{Acknowledgments}
The authors would like to thank an anonymous referee for useful comments and 
suggestions.
This research is co-financed by Greece and the European Union (European Social Fund-ESF) 
through the Operational Programme ``Human Resources Development, Education and Lifelong 
Learning'' in the context of the project ``Strengthening Human Resources Research 
Potential via Doctorate Research'' (MIS-5000432), implemented by the State Scholarships 
Foundation (IKY). This article is based upon work from COST Action ``Cosmology and 
Astrophysics Network for
Theoretical Advances and Training Actions'', supported by COST (European Cooperation in
Science and Technology).

\end{document}